\documentclass[]{imsart}

\usepackage{graphicx,color}
\usepackage{subcaption}
\usepackage{amssymb}
\usepackage{amsmath}
\usepackage{natbib}

\begin{document}

\begin{frontmatter}

\title{Benefits of spatio-temporal modelling for short term wind power forecasting at both individual and aggregated levels} 
\runtitle{Spatio-temporal modelling for short term wind power forecasting}


\begin{aug}
\author{\fnms{Amanda} \snm{Lenzi}\thanksref{t1}\ead[label=eA]{amle@dtu.dk}},
\author{\fnms{Ingelin} \snm{Steinsland}\thanksref{t3}\ead[label=eI]{ingelin.steinsland@ntnu.no}}
\and
\author{\fnms{Pierre} \snm{Pinson}\thanksref{t2}\ead[label=eP]{ppin@elektro.dtu.dk}},


\thankstext{t1}{Research supported by CAPES}
\thankstext{t3}{Research supported by the Research Council of Norway}
\thankstext{t2}{Research supported by the Danish Strategic Council for Strategic Research through the project 5s-Future Electricity Markets}
\runauthor{Lenzi, A. et al.}

\affiliation{Technical University of Denmark\thanksmark{m11} , Norwegian University of Science and Technology \thanksmark{m2}}

\address{Technical University of Denmark\thanksmark{t1}\\
Applied Mathematics and Computer Science Department\\
2800 Kgs. Lyngby Denmark\\
\printead{eA}}

\address{Norwegian University of Science and Technology\thanksmark{t3}\\
Department of mathematical sciences\\
N-7491 Trondheim, Norway\\
\printead{eI}}

\address{Technical University of Denmark\thanksmark{t2}\\
Electrical Engineering Department\\
2800 Kgs. Lyngby, Denmark \\
\printead{eP} }

\end{aug}

\begin{abstract}
The share of wind energy in total installed power capacity has grown rapidly in recent years around the world.
Producing accurate and reliable forecasts of wind power production, together with a quantification of the uncertainty, is essential to optimally integrate wind energy into power systems. 
We build spatio-temporal models for wind power generation and obtain full probabilistic forecasts from 15 minutes to 5 hours ahead. Detailed analysis of the forecast performances on the individual wind farms and aggregated wind power are provided. 
We show that it is possible to improve the results of forecasting aggregated wind power by utilizing spatio-temporal correlations among individual wind farms. Furthermore, spatio-temporal models have the advantage of being able to produce spatially out-of-sample forecasts. We evaluate the predictions on a data set from wind farms in western Denmark and compare the spatio-temporal model with an autoregressive model containing a common autoregressive parameter for all wind farms, identifying the specific cases when it is important to have a spatio-temporal model instead of a temporal one. This case study demonstrates that it is possible to obtain fast and accurate forecasts of wind power generation at wind farms where data is available, but also at a larger portfolio including wind farms at new locations. 

The results and the methodologies are relevant for wind power forecasts across the globe as well as for spatial-temporal modelling in general.

\end{abstract}




\end{frontmatter}

\section{Introduction}
\label{intro}
Wind power is a clean, renewable and widely available source of energy and electricity generated from wind power is increasing world wide. 
A challenge for utilizing wind power is that the generated amount of energy varies much and relatively fast over time due to variations in wind. 
An important tool for efficiently integrating wind power in a system with energy sources that can be controlled, e.g. thermal energy and hydro power, is high quality probabilistic forecasts for short term wind power production \cite{ackermann2005wind}. Moreover, accurate forecasting of wind power generation makes wind more competitive in the energy market, since it reduces the imbalance costs to producers \cite{girard2013assessment}.
Recently, there has been an increasing amount of research in wind speed and wind power forecasts. 
Most of the developments are for point forecasts (e.g. \cite{louka2008improvements}, \cite{catalao2011short}), i.e. the forecast consists of one value for each wind farm or location. 
To make better decisions one also needs to quantify the uncertainty of the forecast, and provide a probability density function (pdf) instead of a point forecast. This is called a probabilistic forecast. 
For a probabilistic forecast to be useful it needs to be calibrated and sharp \cite{gneiting2007probabilistic}.
Calibrated refers to a forecast that is reliable: in the long term, 90$\%$ of the observed wind production should be within a 90$\%$ forecast interval, 80$\%$ of the observations within a 80$\%$ forecast interval and so forth. 
Sharpness refers to the spread of the predictive distribution, a sharper forecast is more concentrated and better when subject to calibration. 

In recent years, more emphasis has been placed on probabilistic forecasts in order to quantify the inherent uncertainties in wind, see \cite{pinson2010conditional} and \cite{bremnes2004probabilistic}. From the point of view of a wind farm operator, probabilistic forecasts improve decision making regarding the management of the immediate regulating and spinning reserves, which is essential given the financial penalties that are incurred for deviating from the declared power levels. 
From the point of view of a system operator, the aggregated wind power generation over pre-defined areas is of particular importance. Some recent contributions to the modelling and forecasting of aggregated wind power energy are \cite{lau2010approaches} and \cite{focken2002short}, which do, however, not account for spatio-temporal dependencies. 

To illustrate the challenge of forecasting individual and aggregated wind power simultaneously, we consider a toy example of two wind farms at
one lead time and denote their forecasts $X_1$ and $X_2$ (these are random variables). 
The aggregated forecast for the system is $Y=X_1 + X_2$. We know from basic probability, see e.g \cite{ross2015first}, that the expected value for the system
is ${\mbox{E}}(Y)={\mbox{E}}(X_1) + {\mbox{E}}(X_2)$ and the variance is ${\mbox{Var}}(Y)= {\mbox{Var}}(X_1) + {\mbox{Var}}(X_2) + 2{\mbox{Cov}}(X_1, X_2)$. 
Hence, to obtain a forecast for the system $Y$ we also need to model the dependency between the wind farms. This calls for a spatio-temporal model for wind power production. 
If the productions at the two farms in our toy example are dependent and have a positive covariance, but are assumed independent in the forecast, the 
variance of $Y$ gets too small and the forecast for $Y$ is not calibrated. Verification of multivariate probabilistic forecasts is an active field of research, for which new scores and diagnostic tools are being proposed and discussed, see, e.g., \cite{pinson2013discrimination}, \cite{scheuerer2015variogram}, \cite{thorarinsdottir2016assessing} among others. A pragmatic approach is to evaluate relevant univariate probabilistic forecasts derived from the multivariate probabilistic forecast. 

Wind speed, and hence wind power production, has temporal and spatial dependencies. In Section \ref{data} we will see that this is also the case for western Denmark. Indeed, our approach of basing the forecast on recent observations relies on the temporal dependency. 
As demonstrated with our toy example, the spatial dependency also needs to be considered for the aggregated forecasts to be calibrated.
Furthermore, borrowing information by utilizing the spatial correlation among individual wind farms has been shown to reduce the errors in point forecasts significantly \cite{tastu2011spatio}, and has the advantage of producing models that are able to generate forecasts at locations that are not within the observation samples.

Several characteristics in a typical wind power series make it a challenging problem to generate accurate forecasts.
First of all, wind power is bounded below by zero, when no turbines are operating, and above by the nominal capacity, when all turbines are generating their rated power output. In addition, wind power series are clearly non-Gaussian. In fact, the marginal distribution of wind power production data possesses tails that are heavier than the Gaussian distribution. Instead of using a classical Gaussian distribution, truncated Gaussian, censored Gaussian and generalized logit-normal distributions have been proposed to model the conditional density of wind power \cite{gneiting2006calibrated} \cite{pinson2012very}. Our approach is based on the logistic function, which has shown to be a suitable transformation to normalize wind power data \cite{dowell2016very}.

We propose statistical models that yield calibrated probabilistic forecasts of wind power generation at multiple sites and lead times simultaneously. 
We define three different models that share the same data process, or likelihood, but differ in the process model. 
We start with a model consisting of a location specific intercept and an autoregressive component that captures the local variability without considering the dependency between the farms. 
This model is well suited for individual forecasts, but it is not calibrated for aggregated forecasts.
To obtain reliable aggregated forecasts, we introduce two different models that capture the spatio-temporal features present in the data. The first has a common intercept and a spatio-temporal process, in which spatial and temporal dependency is modelled by a latent Gaussian field. The second is a combination of the previous two models, with a common intercept, an autoregressive process and a spatio-temporal term that varies in time with first order autoregressive dynamics. 
To meet the computational requirements the stochastic partial differential equations (SPDE) approach to spatial and temporal-spatial modelling is taken \cite{lindgren2011explicit} \cite{blangiardo2015spatial}, for which fast Bayesian inference can be performed using integrated nested Laplace approximations (INLA).

Moreover, we study the performance of the proposed models in forecasting wind power from individual and aggregated farms under two different scenarios.
In a first stage, we consider out-of-sample forecasts in terms of time, that is, they are obtained for wind farms inside the training set. However, there are situations where not enough data is available for all the wind farms, and even when it is available, the computational load to calculate forecasts for all of them can be very high. In those cases, it is important to have a method of forecasting that is as robust as possible, so that parameters estimated using only part of the portfolio can readily be used to forecast a larger data set, including wind farms at new locations. In such cases, temporal models that require local information for the parameter estimation cannot be used to obtain forecasts. Based on this, in a second stage, we consider spatially out-of-sample forecasts generated by the proposed spatio-temporal models. We develop and evaluate the forecasts for wind power production in western Denmark based on a data set for 349 wind farms with energy production observations every 15 minutes from 2006 to 2012. 

In Section~\ref{data}, we provide a short description of the wind power data that we use in our study and the data treatment. The hierarchical models used to generate probabilistic forecasts of wind power generation, as well as the framework for producing probabilistic forecasts with such models, are outlined in Section~\ref{models}. In Section~\ref{evaluation}, we give details of the probabilistic forecasting scheme and outline the scores and the scenarios used for forecast evaluation. In Section~\ref{results}, we show the results of a case study where we obtain spatio-temporal forecasts and spatially out-of-sample forecasts on the individual and aggregated level. Section~\ref{results} also contains the results of a simulation study, whereas conclusions of the work are drawn in Section \ref{conclusions}. 

\section{Danish wind power production data}
\label{data}
This project is based on a system of 349 wind farms in western Denmark. Observations of wind power production between January 2006 and March 2012 were provided by the Transmission System Operator in Denmark and each measurement consists of temporal average over a 15-min time period.

The measurements at each site have been normalised by the nominal power of the corresponding wind farm, so that they are within the range [0,1]. Moreover, to avoid including long chains of zeros that come from temporary shutdown of the turbines for maintenance or missing data that are reflected as unreasonably long periods of zero wind power production, we choose to analyze only wind farms containing at most $10\%$ of zero observations. 
The evaluation of the predictive performance of individual wind farms and aggregated wind power is done as $\%$ of nominal power, which is a common practice in the wind power field (e.g., \cite{pinson2012very}, \cite{tastu2011spatio}, \cite{dowell2016very}).

\begin{figure}[htb!]
\centering
\begin{subfigure}{.5\textwidth}
  \centering
  \includegraphics[width=1.2\linewidth]{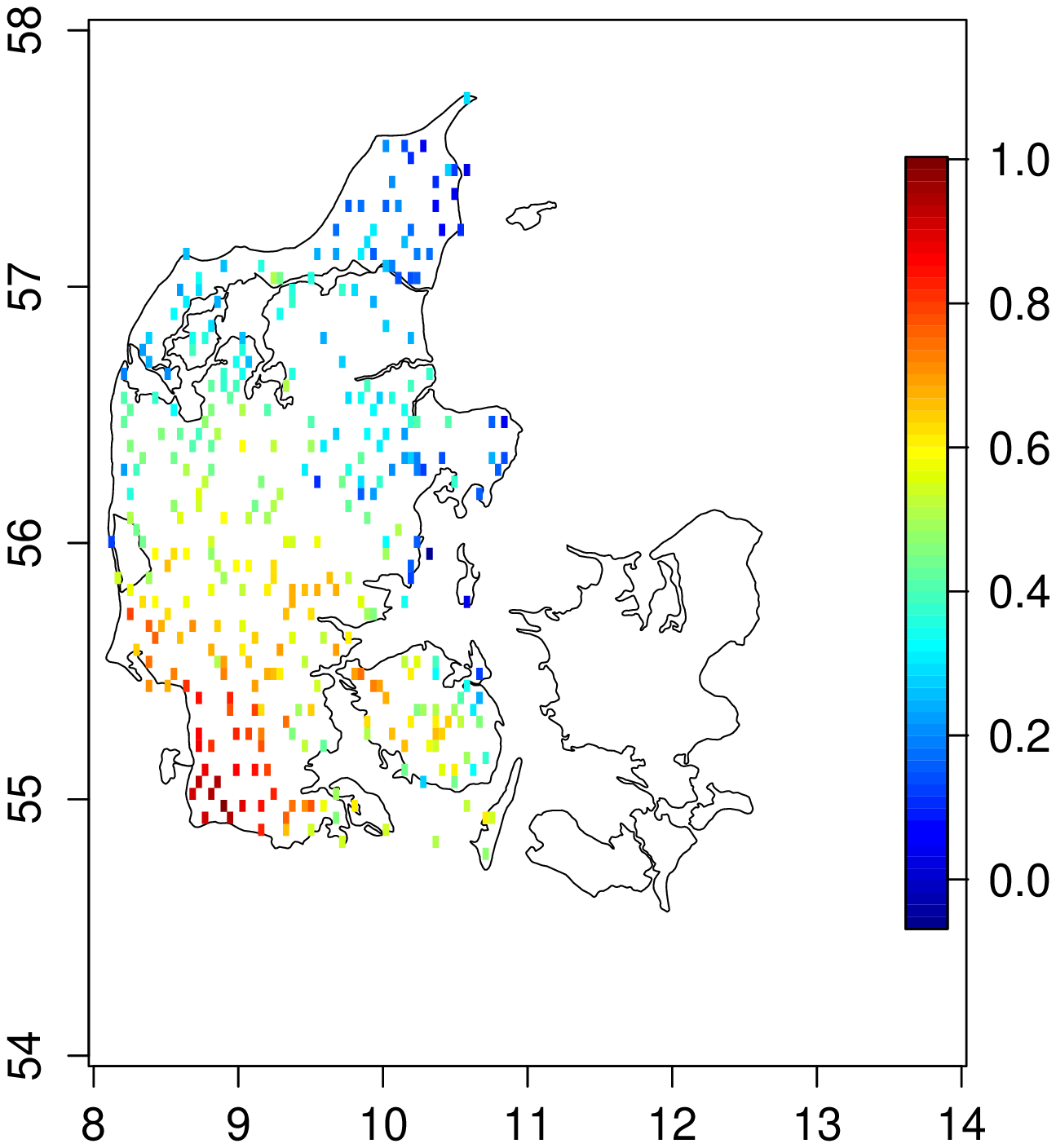}
  \caption{}
  \label{fig:sub1}
\end{subfigure}%
\begin{subfigure}{.5\textwidth}
  \centering
  \includegraphics[width=0.9\linewidth]{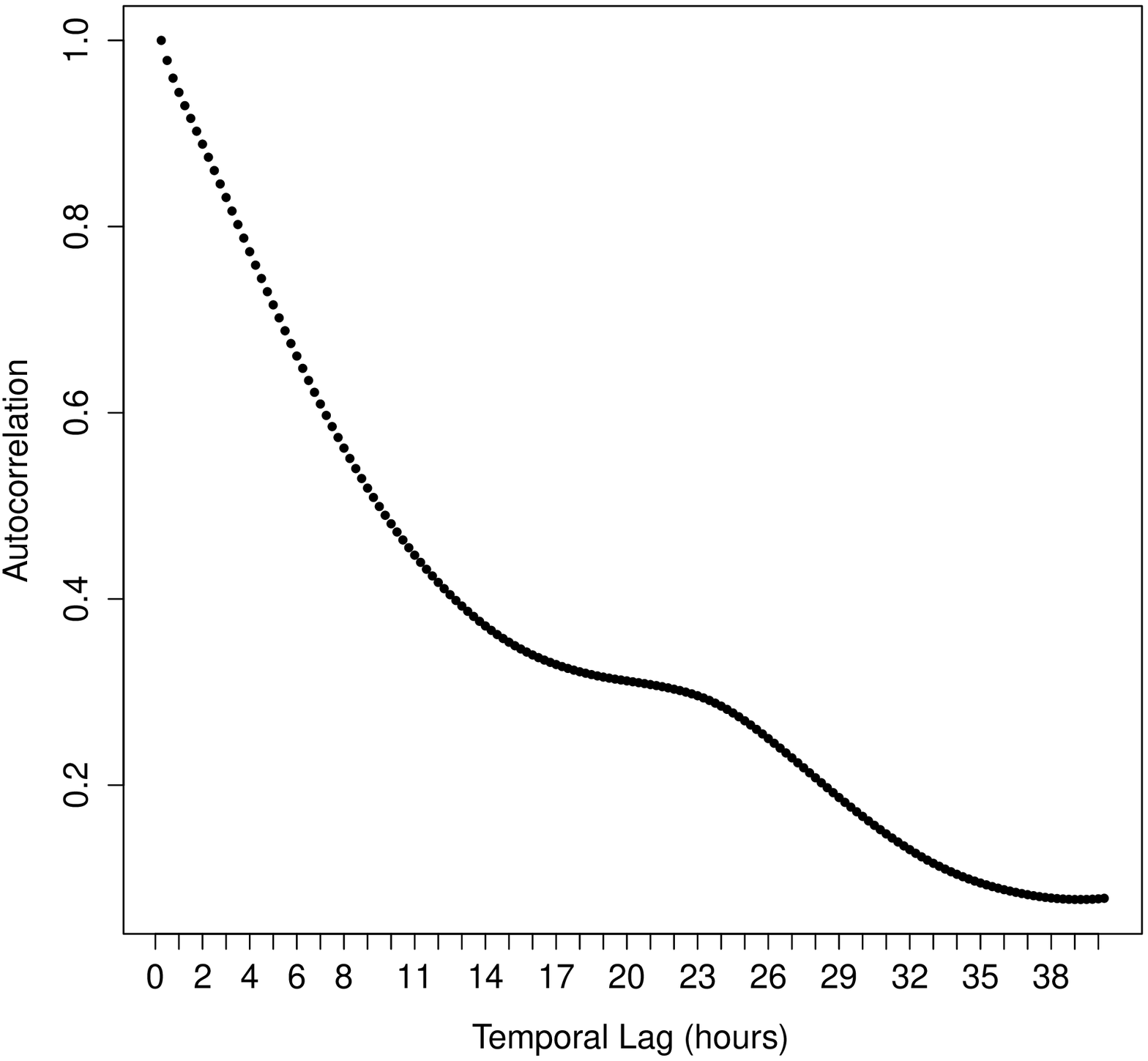}
  \caption{}
  \label{fig:sub2}
\end{subfigure}
\caption{ (a) Map of spatial correlation of wind power production between one wind farm located in the southern part of western Denmark and the remaining wind farms. The correlations between wind farms in a closer proximity are clearly higher than between wind farms that are farther apart. (b) Mean autocorrelation function of wind power production at wind farms located in western Denmark. The autocorrelations decay slowly.}
\label{fig:map-acf}
\end{figure}

Figure \ref{fig:map-acf} (a) shows the spatial correlation of wind power production between one wind farm located in the southern part of Denmark and the remaining wind farms of the portfolio. The higher correlations come from farms that are closer, while the correlations of wind farms far from it are almost zero. Next, we check the dependency of the temporal correlation at fixed locations. Figure \ref{fig:map-acf} (b) shows the mean autocorrelation function of wind power production among wind farms located in western Denmark. The autocorrelation function of the normalized wind power production at a single farm has a slow decay and on average, it drops down to zero after about 40 hours. 

Wind power generated by a farm over a period of time is non-Gaussian and bounded between zero and one after the normalization. In fact, wind power distribution has a sharper peak than the Gaussian distribution and is also significantly right-skewed.
In all the approaches to be described next, we apply the logit-normal transformation to the normalized wind power data following the procedure in \cite{pinson2012very}. 


Let $X({\mathbf{s}},t)$ denote the normalized wind power production at location ${\mathbf{s}} \in \mathcal{D}_s$ and time $t \in \mathcal{D}_t$, with respective observations or measurements indicated by $x({\mathbf{s}},t)$.
The logit-normal transformation is given by
\begin{equation}
y({\mathbf{s}},t) = \gamma(x({\mathbf{s}},t)) = {\mbox{ln}} \Big( \frac{x({\mathbf{s}},t)}{1-x({\mathbf{s}},t)} \Big), \quad x({\mathbf{s}},t) \in (0,1),
\label{logit-trans}
\end{equation}
with inverse
\begin{equation}
x({\mathbf{s}},t) = \gamma^{-1}(y({\mathbf{s}},t)) = (1 + e^{-y({\mathbf{s}},t)})^{-1}, \quad y({\mathbf{s}},t) \in \mathbb{R}.
\label{logit-inv-trans}
\end{equation}
To represent the logit-normal transformation in the cases where measurements are equal to zero and one, we follow the approach by \cite{lesaffre2007logistic} for modelling outcome scores in $[0, 1]$.

Moreover, to evaluate the performance of aggregated wind power forecasts, we obtain the normalized aggregated wind power at lead time $h$ by
\begin{equation}
 x_A(t+h) = \frac{\sum_{j=1}^{N} c_j x({\bf{s}}_{j}, t+h)}{\sum_{j=1}^{N} c_j},
 \label{eq:agg}
\end{equation}
where $c_j$ is the capacity of wind farm at location ${\bf{s}}_{j}$ and $N$ is the total number of wind farms in the portfolio.


\section{Models and fitting scheme}
\label{models} 

In this section, we introduce three different statistical models for wind power production. 
We start with a simpler autoregressive model, where each wind farm is considered as an independent replicate of the same process. Next, we describe two versions of a spatio-temporal model, in which spatial correlation is captured by a latent Gaussian field with a Mat\'{e}rn covariance function.
The simplest version has only a spatio-temporal component, while the other has both, an autoregressive process and a spatio-temporal model.
The section ends with the estimation procedure and how we obtain probabilistic forecasts.

\subsection{Likelihood}
\label{likelihood}

We denote by $Y({\mathbf{s}},t)$ the normalized logit-normal transformed wind power generation at location ${\mathbf{s}}$ and time $t$, which is calculated using (\ref{logit-trans}). We assume the following distribution for $Y({\mathbf{s}},t)$ at the first level of the hierarchical models considered in this section
\begin{equation}
Y({\mathbf{s}},t) \sim \mbox{Normal}\left(\mu({\mathbf{s}},t), \sigma^2_e \right),
\label{norm}
\end{equation}
with $\sigma^2_e$ being the variance of the measurement error, defined by a Gaussian white noise process both serially and spatially uncorrelated. The
term $\mu({\mathbf{s}},t)$ is the mean of the random process and can be defined by other process levels giving rise to different hierarchical models that are described in the following sections.


\subsection{Latent Gaussian structure}
\label{latent}

\subsubsection{Temporal model (Model T)}
\label{temporal}

We start with a time series model where each wind farm is considered as an independent replicate of the same random process. The independence assumption is of course a simplification, since the wind power production in one location is probably dependent on the production in other locations. 
We assume that $\mu({\mathbf{s}},t)$, in (\ref{norm}), is constant in time and can be modelled as
\begin{equation}
\mu(\mathbf{s},t) = b(\mathbf{s}) + w_{\mathbf{s}}(t), 
\label{model1}
\end{equation}
where $ b(\mathbf{s})$ is an intercept specific for each location and $w_{\mathbf{s}}(t)$ is an autoregressive process that can be written as
\begin{equation}
w_{\mathbf{s}}(t) = \rho_1 w_{\mathbf{s}}(t-1) + \nu_{\mathbf{s}}(t),
\label{ar1}
\end{equation}
with $t = 2, \ldots, T$ and $|\rho_1| < 1$. The term $ \nu_{\mathbf{s}}$ is uncorrelated with $w_{\mathbf{s}}(t)$ and independent identically distributed as $\nu_{\mathbf{s}} \sim N(0, \sigma^2_{\nu})$.

\subsubsection{Spatio-temporal model  (Model S-T)}
\label{spatio-temporal}

This model is a spatio-temporal process with temporal dynamics as in \cite{cameletti2013spatio}. This type of model is commonly used for modelling air quality because of its flexibility in including time and space dependency, as well as the effect of covariates (see e.g. \cite{fasso2011maximum} and \cite{cocchi2007hierarchical}).
The mean function $\mu({\mathbf{s}},t)$ in (\ref{norm}) is given by 
\begin{equation}
\mu({\mathbf{s}},t) = b_0 + z({\mathbf{s}},t),
\end{equation}
where $b_0$ is an intercept that is common to all wind farms and constant in time and space. The term $z({\mathbf{s}},t)$ refers to a spatio-temporal process that varies in time with first order autoregressive dynamics  
\begin{equation}
z({\mathbf{s}}, t) = \rho_2 z({\mathbf{s}}, t-1) + w({\mathbf{s}}, t),
\label{spatiotemporal}
\end{equation}
with $t = 2, \ldots, T$ and $|\rho_2| < 1$. Moreover, $w(s, t)$ is a zero-mean Gaussian field, assumed to be temporally independent with covariance function
\[
    {\mbox{Cov}}(w({\mathbf{s}}, t), w({\mathbf{s}}', t') ) = 
\begin{cases}
    \sigma_w^2C(h),  & \text{if } t =  t' \\        0,         & t \neq t'
\end{cases}
\]
for $\mathbf{s} \neq \mathbf{s}'$. The correlation function $C$ depends on the locations ${\mathbf{s}}$ and ${\mathbf{s}}'$ through the distance $h=||{\mathbf{s}} - {\mathbf{s}}'||$. This means that the process is assumed to be second-order stationary and isotropic (see \cite{cressie1992statistics}). The marginal variance is $\mbox{Var}({\mathbf{s}}, t) = \sigma_w^2$ and $C(h)$ is the correlation function defined by the Mat\'{e}rn, given by

\begin{equation}
C(h) = \frac{1}{\Gamma(\nu)2^{\nu-1}}(\kappa h)^{\nu} K_\nu(\kappa h),
\label{matern}
\end{equation}
where $K_1$ is the modified Bessel function of second kind, order $\nu$. The parameter $\kappa$ can be used to select the range, while $\nu$ is a smoothness parameter determining the mean-square differentiability of the underlying process. More precisely, the range is defined to be $r = \sqrt{8 \nu}/\kappa$. Although the parameter $\nu$ is fixed to 1 for computational reasons, it remains flexible enough to handle a broad class of spatial variation \cite{rue2009approximate}. Applications with fixed parameter $\nu$ include \cite{ingebrigtsen2014spatial}, \cite{cameletti2013spatio} and \cite{munoz2013estimation}.


\subsubsection{Temporal + Spatio-temporal model (Model ST+T)}
\label{temp+spatio-temporal}

This is a model defined by an autoregressive process at each location to capture the individual variability and a spatio-temporal process with temporal dynamics to take into account the spatial dependence among wind farms. Specifically, $\mu({\mathbf{s}},t)$  from (\ref{norm}) is defined as
\begin{equation}
\mu({\mathbf{s}},t) = b_0 +  w_{\mathbf{s}}(t) + z({\mathbf{s}},t),
\end{equation}
where $b_0$ is a fixed unknown intercept that is shared by all wind farms. The process  $w_{\mathbf{s}}(t)$ is assumed to have autoregressive dynamics as defined in (\ref{ar1}). Finally,  $z({\mathbf{s}},t)$ is a spatio-temporal component that has the structure of (\ref{spatiotemporal}) and its spatio-temporal covariance function is the same as in (\ref{matern}).


For all the models described above, a log-Gamma prior is assumed for the parameters in the  Mat\'{e}rn covariance as well as for the precision parameters $\sigma^2_e$ and $\sigma_\nu^2$. For the fixed effect $b$'s we assume Gaussian priors. The correlations $\rho$'s are specified over the parametrization $\mbox{log}(\frac{1+\rho}{1-\rho})$ with prior Gaussian distributions.

\subsection{Inference and prediction} 
\label{background}

The key feature of the models described above is that they can be handled within the theoretical and computational framework developed by \cite{rue2009approximate} and \cite{lindgren2011explicit}. The approach by \cite{rue2009approximate} allows us to directly compute accurate and fast approximations of the posterior marginals. In addition, the method by \cite{lindgren2011explicit} is computationally efficient for inferential purposes: instead of using a Gaussian random fields (GRF) with dense covariance matrix, the computations are carried out with a  Gaussian Markov random field (GMRF) with sparse precision matrix. 
The original idea comes from the work of \cite{whittle1954stationary} and \cite{whittle1963stochastic}, where it is shown that the solution to the SPDE 
\begin{equation}
(\kappa^2 - \Delta)^{\alpha/2} x({\bf{u}}) = \mathcal{W}({\bf{u}}), \quad  {\bf{u}} \in \mathbb{R}^d, \alpha = \nu + D/2, \kappa>0, \nu>0,
\label{eq:spde}
\end{equation} 
is a GRF with Mat{\'e}rn covariance function. The innovation process $\mathcal{W}$ on the right hand side of (\ref{eq:spde}) is Gaussian white noise and $\Delta$ is the Laplacian. 

\begin{figure}[htb!]
\centering
  \includegraphics[width=0.5\textwidth]{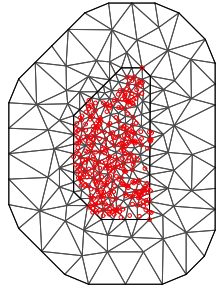}
\caption{The western Denmark triangulation. The red dots denote the observation locations of the wind power production data.}
\label{mesh}      
\end{figure}

An approximation to the solution of the SPDE in (\ref{eq:spde}) can be obtained using the finite element method (FEM), a numerical technique for solving partial differential equations \cite{lindgren2011explicit}. This is done by representing the infinite dimensional GRF by a linear combination of finite basis function
\begin{equation}
x({\bf{u}}) = \sum_k{\psi_k({\bf{u}})}w_k
\label{eq:triang}
\end{equation} 
where the $w_k$'s are random weights chosen so that the representation in (\ref{eq:triang}) approximates the distribution of the solution to the SPDE in (\ref{eq:spde}). 
The $\psi_k$'s are basis functions defined on a triangulation of the domain, i.e. a subdivision into non-intersecting triangles. Figure \ref{mesh} shows the triangulation of western Denmark data set described in Section \ref{data}.

Next, the posterior estimates of parameters and hyperparameters are computed using INLA \cite{rue2009approximate}. This method approximates the integral involved in the calculation of the marginal posterior distributions of the hyperparameters by Laplace approximation, making use of the Markov structure of the latent variables in the computation.  
We use the R-INLA package to perform inference and prediction. For more information on the package see http://www.r-inla.org.


\section{Forecast evaluation}
\label{evaluation}


\subsection{Probabilistic forecasting scheme}
\label{implement}

We evaluate the predictive performance of the models described in Section~\ref{models}, using a time moving window approach with data from western Denmark in 2009, so that each training set consists of $L = 2\times96=192$ observations, i.e., two days. In total, the model is fit to $364/2 = 182$ different data sets. We obtain forecasts for lead times $h = 1, \ldots,20$, that is, from 15 minutes up to 5 hours following the training data. Notice that we have compared different lengths of data window $L$ with respect to the root mean squared error (RMSE) and continuous ranked probability score (CRPS). The temporal model presented in Section \ref{temporal} is very sensitive to the window length, such that less than two days of observations in the training set resulted in poor estimation at all lead times. On the other hand, the spatio-temporal models in Section \ref{spatio-temporal} and \ref{temp+spatio-temporal} showed to be robust for different values of $L$, with small changes in the forecast performance for different training sets.

Moreover, because of the high-time resolution of the Danish wind power time series (15-minutes) and the dependency structure in space and time of Model S-T and Model ST+T, the fitting can be very computationally expensive. One way to deal with high-time resolution data is to define the model on a set of knots instead of all time points. Knot-based linear combinations are widely used to tackle computational problems in large data sets (e. g. \cite{paciorek2007bayesian} and \cite{wikle1999dimension}). To fit the spatio-temporal component $z({\mathbf{s}},t)$ in (\ref{spatiotemporal}), we define a set of equally spaced knots at every 12 data points (3 hours), such that the points in time are reduced to only 17 knots, instead of the original 192 observations. Note that the component $w_{\mathbf{s}}(t)$ in models Model T and Model ST+T is fitted to the complete training data, since it does not involve spatio-temporal interactions. 

We evaluate probabilistic forecasts of wind power production from individual wind farms and aggregated. 

Let $\hat{X}({\bf{s}}_{j}, t+h)$ denote the random variable of the wind power forecast at wind farm ${\bf{s}}_{j}$ and lead time $h$. The aggregated forecast of wind power generation is taken as 
\begin{equation}
\hat{X}_{A(t+h)} = \frac{\sum_{j=1}^{N} c_j \hat{X}({\bf{s}}_{j}, t+h)}{\sum_{j=1}^{N} c_j},
\end{equation}
where $c_j$ is the capacity of wind farm ${\bf{s}}_{j}$ and $N$ is the number of wind farms. 
To find the pdf of the aggregated forecasts, $\hat{f}_{X_{A(t+h)}}$, the joint distribution for all wind farms $\{ \hat{X}({\bf{s}}_{1}, t+h), \hat{X}({\bf{s}}_{2}, t+h), \ldots \hat{X}({\bf{s}}_{N}, t+h) \}$ needs to be assessed. Finally, point forecast of aggregated wind power production is obtained as the mean (or median) of $\hat{f}_{X_{A(t+h)}}$.

\subsection{Point and probabilistic forecast scores}
\label{scores}

We assess the quality of predictive performance of the models proposed in Section \ref{models} using both point and probabilistic forecast scores. We obtain point forecast at a specific location as the mean of the forecast density. For each lead time, point forecast of individual power is assessed using the root mean squared error (RMSE), where the mean is taken over all wind farms and data sets, 
\begin{equation}
\mbox{RMSE}(t+h) = \sqrt{\frac{1}{D N} \sum_{i=1}^{D}  \sum_{j=1}^{N} (x({\bf{s}}_{ij}, t+h) - \hat{x}({\bf{s}}_{ij}, t+h))^2  }
\label{rmse_individual}
\end{equation} 
where $D$ is the number of data sets, $N$ is the number of wind farms and $\hat{x}({\bf{s}}_{ij}, t+h) = \gamma^{-1}( \hat{y}({\bf{s}}_{ij}, t+h))$ is the predicted value of $ x({\bf{s}}_{ij}, t+h)$.

To evaluate the performance of forecast densities, we use the continuous ranked probability score (CRPS). \cite{gneiting2007strictly}  showed that CRPS is a strictly proper scoring rule for the evaluation of probabilistic forecasts of a univariate quantity that assesses calibration and sharpness simultaneously \cite{gneiting2007strictly}. A lower score indicates a better density forecast. It is defined as

\begin{equation}
\mbox{CRPS}(F, x) = \int^{\infty}_{-\infty} (F(y) - \delta_{\{y\ge  x\}})^2 dy
\end{equation}
where $F$ is the cumulative distribution function of the density forecast and $y$ is the observation. With the available samples, we can approximate the mean CRPS at each lead time by
\begin{equation}
\begin{aligned}
\label{eq:crps}
\mbox{CRPS}_{F, x}(t + h) = &\frac{1}{D N} \sum_{i=1}^{D}  \sum_{j=1}^{N} \Big(\frac{1}{n}  \sum_{k=1}^{n} |\hat{x}^{(k)}(\mathbf s_{ij},t+h) -x(\mathbf s_{ij},t+h)| \\
 & - \frac{1}{2n^2} \sum_{k,l=1}^{n} |\hat{x}^{(k)}(\mathbf s_{ij},t+h) -\hat{x}^{(l)}(\mathbf s_{ij},t+h)| \Big) ,
\end{aligned}
\end{equation}
where n is the number of samples. Again, the mean CRPS is taken over all the wind farms and data sets in the training set.

Reliability, also referred to as calibration, of probabilistic forecasts is assessed with reliability diagrams. In a calibrated forecast, the observed levels should match the nominal levels for specific quantile forecasts, which results in points aligning with the diagonal in the reliability diagram. To construct reliability diagrams, we start by introducing an indicator variable $\mathcal{I}^{(\alpha)}(\mathbf{s}_{ij}, h)$, which is defined for a quantile forecast $\hat{q}^{(\alpha)}(\mathbf{s}_{ij}, t+h)$ issued at lead time $h$ and wind farm $\mathbf{s}_i$ of the training data $j$, with observed value $x(\mathbf{s}_{ij}, t+h)$ as follows
\[ \mathcal{I}^{(\alpha)}(\mathbf{s}_{ij}, h) =
  \begin{cases}
    1       & \quad \text{if } \quad x(\mathbf{s}_{ij}, t+h) \le \hat{q}^{(\alpha)}(\mathbf{s}_{ij}, t+h) \\
    0,  & \quad {\mbox{otherwise}}\\
  \end{cases}
\]
The indicator variable $\mathcal{I}^{(\alpha)}(\mathbf{s}_{ij}, h)$ shows whether the actual outcome lies below the $\alpha$ quantile forecast (hits) or not (miss). 
Next, $n_{h,1}^{(\alpha)}$ denotes the sum of hits and $n_{h,0}^{(\alpha)}$ the sum of misses over all the realizations
\[
n_{h,1}^{(\alpha)} = \sum^{D}_{i=1} \sum^{N}_{j=1} \mathcal{I}^{(\alpha)}(\mathbf{s}_{ij}, h) \quad \mbox{and} \quad n_{h,0}^{(\alpha)} = DN - n_{h,1}^{(\alpha)}.
\]
An estimation ${\hat{a}}^{(\alpha)}_{h}$ of the actual coverage $a^{(\alpha)}_{h}$ is then obtained by calculating the mean of $\mathcal{I}^{(\alpha)}(\mathbf{s}_{ij}, h)$ over the $N$ wind farms in the $D$ validation sets
\begin{equation}
{\hat{a}}^{(\alpha)}_{h} = \frac{1}{DN} \sum^{D}_{i=1} \sum^{N}_{j=1} \mathcal{I}^{(\alpha)}(\mathbf{s}_{ij}, h) = \frac{n_{h,1}^{(\alpha)}}{n_{h,1}^{(\alpha)} + n_{h,0}^{(\alpha)}}.
\end{equation}
Here, we use nominal levels from 5\% to 95\% in steps of 5\%. Since the number of observations used to calculate the reliability diagrams is of limited size and the observed proportions are equal to the nominal ones only asymptotically \cite{toth2003forecast} \cite{brocker2007scoring}, we follow the idea of \cite{brocker2007scoring} of generating consistency bars for reliability diagrams. 

\subsection{Evaluation scheme}
\label{evaluation-scheme}

We evaluate probabilistic forecasts of Danish wind power production from two different scenarios. First, we consider time forward forecast performances at the locations of the training set. 
The spatio-temporal models, i.e, Model S-T and Model ST+T, have the advantage of being able to provide forecasts where recent observations are not available. Based on this, in a second evaluation scheme, we study the performances of spatially out-of-sample forecasts, which are based on $k$-fold cross-validation with $k=5$. Notice that overall, 5 to 10-fold cross-validation is recommended as a good compromise between bias and variance (\cite{breiman1992submodel}; \cite{kohavi1995study}). The forecast performance measures from the second scenario are obtained by combining the estimates from the 182 data sets in the training set. 

Finally, we validate our results with a simulation study consisting of 200 simulated spatio-temporal data sets. In each data set, logit transformed wind power production measurements, $y({\bf{s}}_{i}, t)$, are "observed" at 200 wind farms belonging to the wind power data set (see left plot of Figure \ref{fig:map-acf}). To mimic the case study based on the Danish wind data set, we simulate data every 15 minutes for 2 days and 5 hours. In total, there are $2 \times 96 + 20 = 212$ measurements taken at each location. All data sets are generated according to Model ST+T directly using the SPDE model construction. 
We use the set of parameters found for one specific data set of the training set from fitting Model ST+T to the logit transformed Danish wind power data. 


\section{Results}
\label{results}

In this section we show the results from a case study, where we use the models described in Section~\ref{models} to forecast individual and aggregated wind power in Denmark. As described in Section~\ref{evaluation-scheme}, we evaluate and discuss the performances of our models when we consider time forward forecasts at the locations of the training set. We call these spatio-temporal forecasts, and we also show the case of spatially out-of-sample forecasts, i.e, for wind farms that are not in the training set. Furthermore, we illustrate the results from a simulation study based on our case study. Details of the probabilistic forecasting scheme can be found in Section~\ref{implement}, while the methodology used to rank point and probabilistic forecasts is in Section~\ref{scores}.

\subsection{Spatio-temporal forecast performance} 
\label{insample}

\begin{figure}[!htb]
\centering
\begin{subfigure}{0.9\textwidth}
\centering
  \includegraphics[width=1\linewidth]{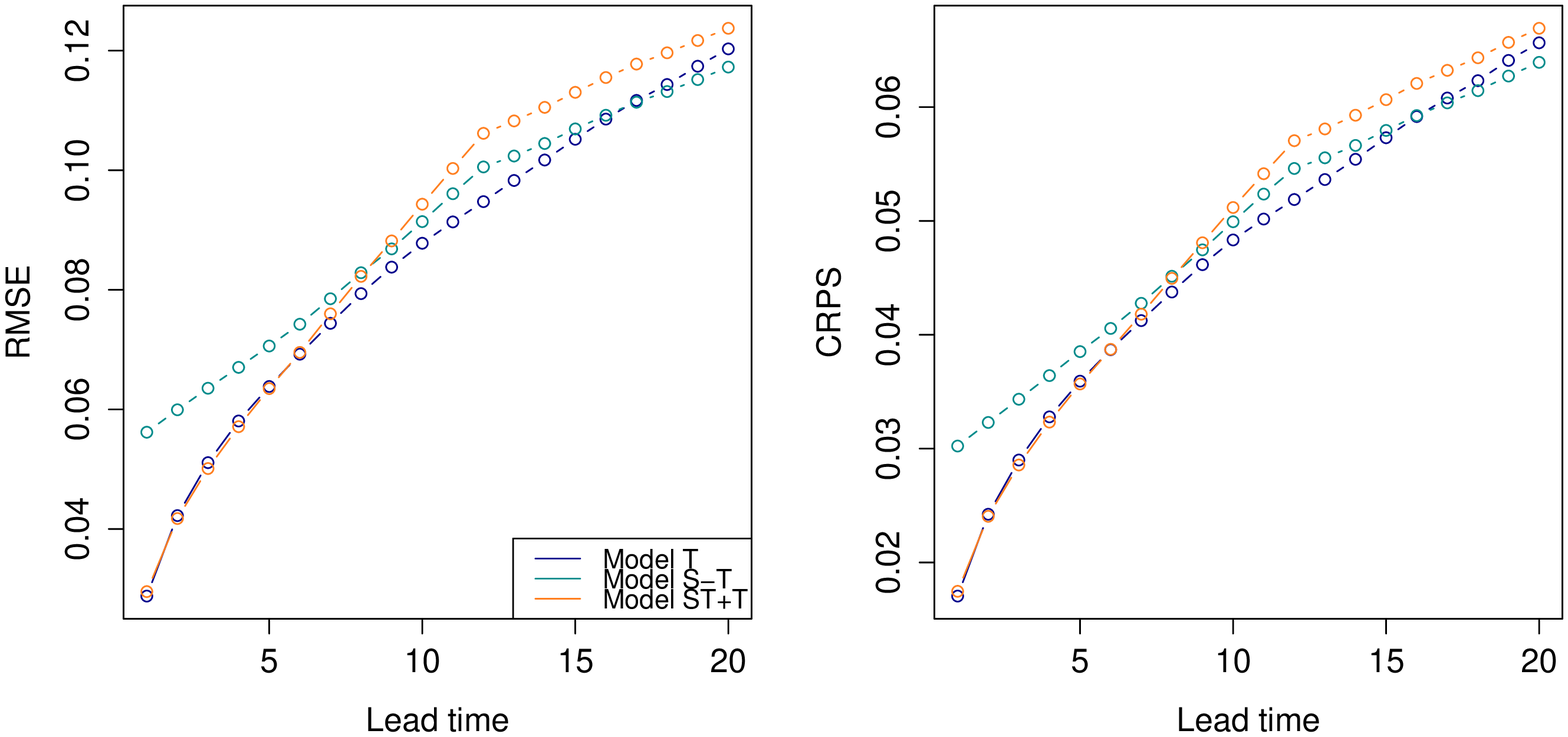}
    \caption{}
\end{subfigure}%

\begin{subfigure}{0.9\textwidth}
\centering
  \includegraphics[width=1\linewidth]{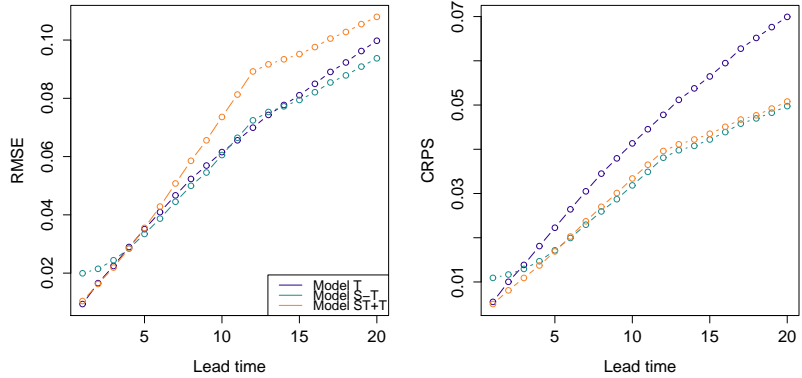}
    \caption{}
\end{subfigure}
\caption{RMSE and CRPS (as \% of nominal power) of spatio-temporal wind power forecasts at lead times $1, \ldots, 20$ (i.e., from 15 minutes up to 5 hours) for Model T (blue), Model S-T (green) and Model ST+T (orange). (a) Forecasts for individual wind farms. (b) Forecasts for aggregated wind farms.}
\label{errors_orig} 
\end{figure}

Figure~\ref{errors_orig} summarizes the spatio-temporal forecast performances of the three models introduced in Section~\ref{models} in terms of RMSE and CRPS. As we can see from Figure~ \ref{errors_orig} (a), Model T and Model ST+T outperformed Model S-T with respect to RMSE and CRPS when forecasting individual wind farms at lead times 1-6 (i.e, from 15 minutes up to 2 hours ahead). For higher lead times, the three models have similar performance. 
In terms of aggregated wind power production, Model T performed similar to Model ST+T in terms of point forecast (RMSE), but it has poor performance according to CRPS values, as shown in Figure~ \ref{errors_orig} (b).

\begin{figure}[!htb]
\centering
\begin{subfigure}{0.9\textwidth}
\centering
  \includegraphics[width=0.8\linewidth]{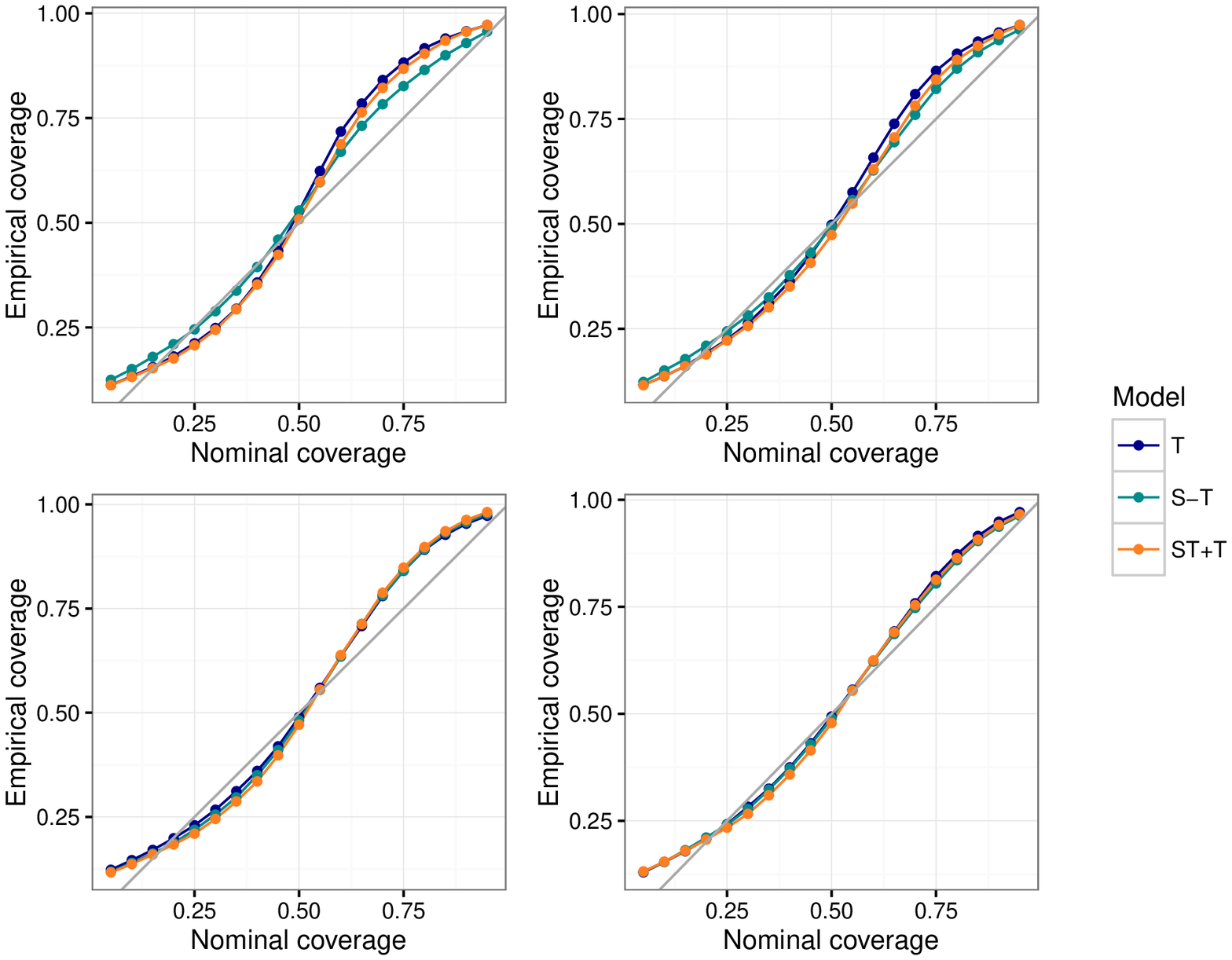}
    \caption{}
\end{subfigure}%

\begin{subfigure}{0.9\textwidth}
\centering
  \includegraphics[width=0.8\linewidth]{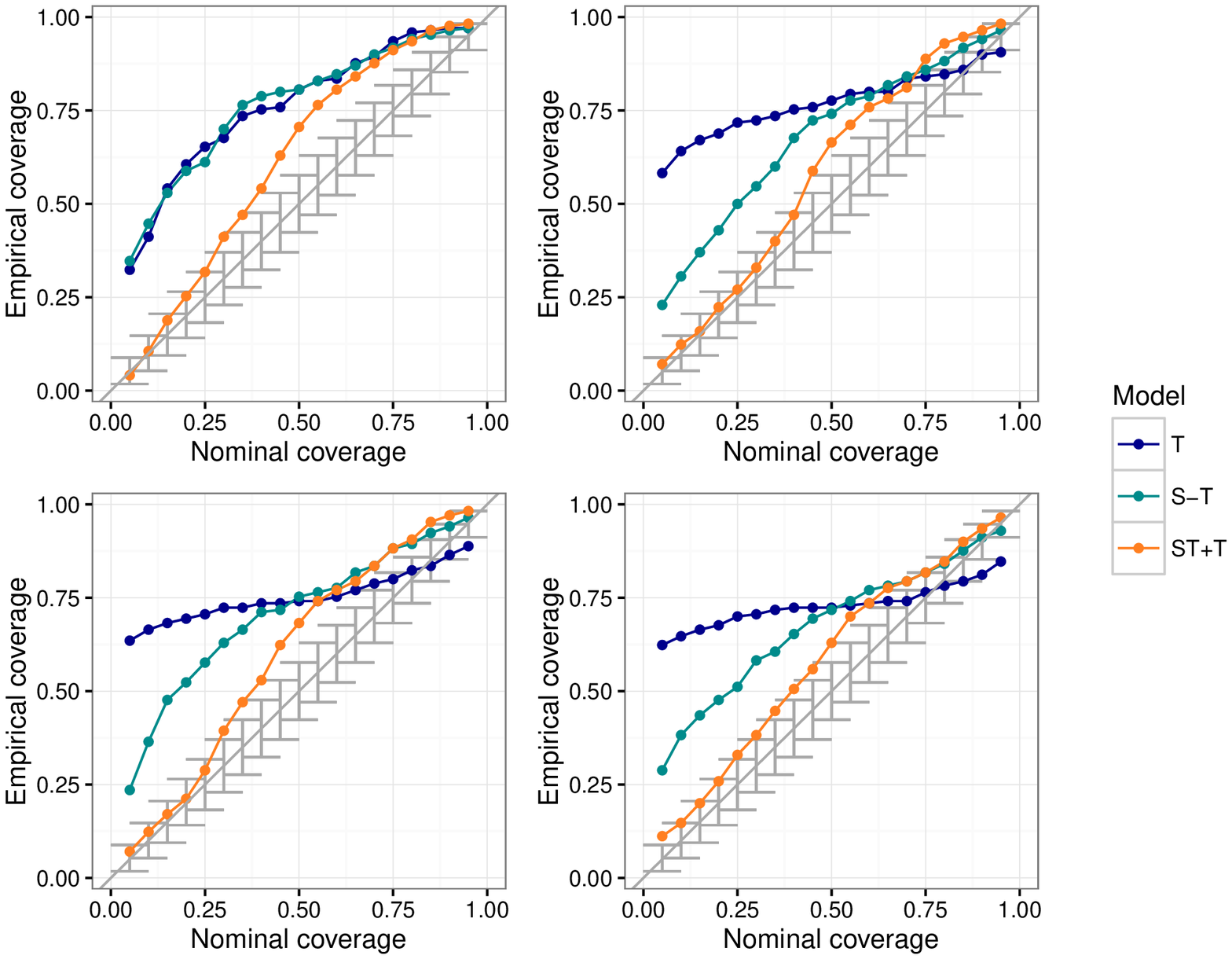}
    \caption{}
\end{subfigure}
\caption{ Reliability diagram of spatio-temporal wind power forecasts at lead time 1 ({\it{Top left}}), 7 ({\it{Top right}}) , 13 ({\it{Bottom left}})  and 19 ({\it{Bottom right}}). The diagrams were calculated using Model T (blue), Model S-T (green) and Model ST+T (orange). (a) Forecasts for individual wind farms. (b) Forecasts for aggregated wind farms.}
\label{reliab_orig}
\end{figure}

Reliability diagrams for each model at lead times $h=1,7,13$ and 19 are presented in Figure \ref{reliab_orig}. These diagrams compare the theoretical and the observed proportions of a set of quantiles from forecasts made at all wind farms and data sets in the training set. The forecasts at individual wind farms produced by the three models presented in Section \ref{models} perform similarly well in terms of reliability, with points close to the diagonal for most quantiles, see Figure \ref{reliab_orig} (a).
Since the number of observations used to calculate reliability diagrams is relatively small (182 data sets in the training set), consistency bars for the evaluation of forecasts from aggregated farms are also plotted, as shown in Figure \ref{reliab_orig} (b). The aggregated forecasts provided by Model ST+T are the best calibrated among the three models for most of the quantiles at all lead times, followed by Model S-T. Even though the performance of Model T is comparable with the performance of the other models in terms of aggregated forecast density mean (RMSE), we can see that this model does not produce reliable probabilistic forecasts for the aggregated data. This fact is more obvious for the lower quantiles; more than 50\% of the observed aggregated forecasts are below the nominal 5\% quantile at lead times $h=7, 13$ and 19. 

\begin{figure}[htb!]
\centering
   \includegraphics[width=1\textwidth]{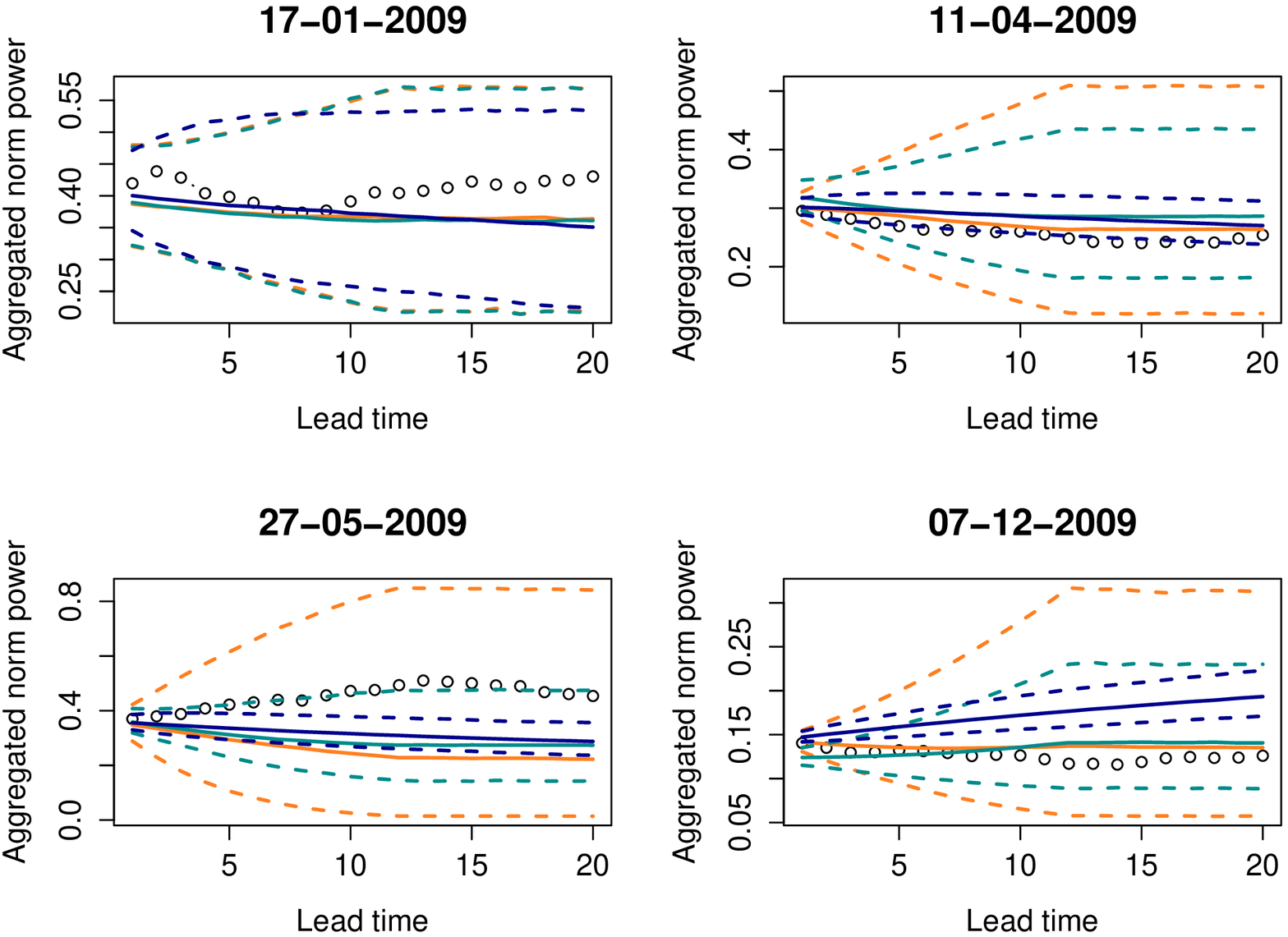}
\caption{$5\%$ and $95\%$ quantiles (dashed lines), as well as the median (solid lines) of the aggregated forecast densities from four different data sets in the training set, together with the actual observed aggregated power produced (circles) at lead times 1-20 (i.e., from 15 minutes up to 5 hours). The forecast densities correspond to Model T (blue), Model S-T (green) and Model ST+T (orange).
An example of a data set where all the models have forecast densities that cover the actual aggregated production is shown in the {\it{Top left}} plot.
In the {\it{Top right}} plot, the observations lie close to the median of the forecast densities from Model S-T and Model ST+T, but close to the $5\%$ quantile of the forecast density from Model T. 
{\it{Bottom left}} and {\it{Bottom right}} plots illustrate cases where Model T has forecast densities that are too narrow and fail to predict the aggregated wind power, while the forecasts from Model ST+T provide densities that are wide enough to cover the true value at all lead times.} 
\label{quants_sum_orig}       
\end{figure}

We further explore aggregated probabilistic forecasts from models in Section \ref{models} with plots containing the $5\%$, $50\%$ and $95\%$ quantiles of the aggregated forecast densities together with the actual observed aggregated power produced at four different data sets in the training set, as shown in Figure \ref{quants_sum_orig}. We noticed that Model T results in forecast densities that are consistently too narrow.
On the other hand, Model ST+T provides the widest aggregated forecast densities among the three models in most of the data sets, which produces calibrated forecasts at all lead times. This is confirmed in Figure~\ref{reliab_orig} (b) and will be further explored in the simulation studies in Section~\ref{simulation}. 

\subsection{Spatially out-of-sample forecast performance} 
\label{outofsample}
\begin{figure}[!htb]
\centering
\begin{subfigure}{0.9\textwidth}
\centering
  \includegraphics[width=1\linewidth]{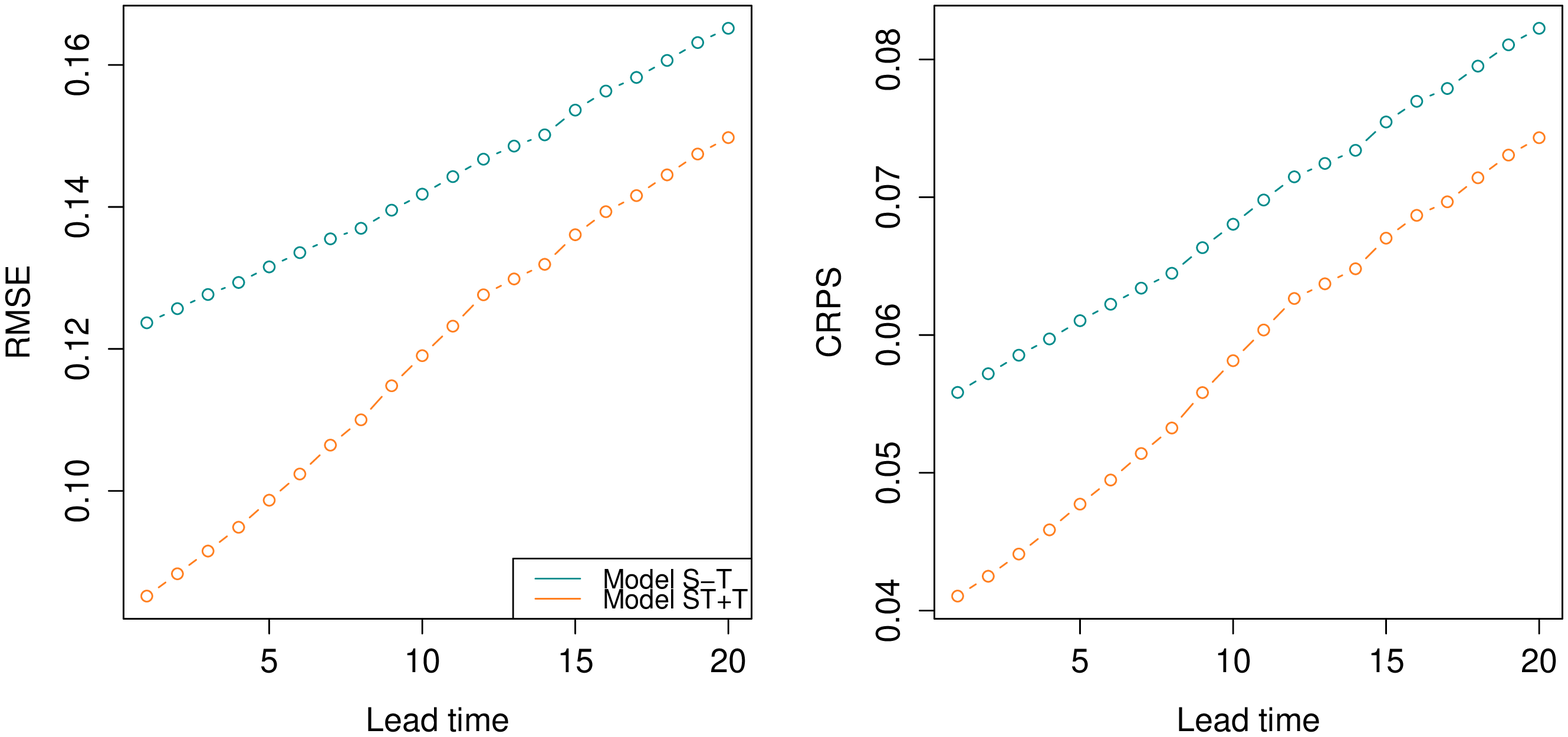}
    \caption{}
\end{subfigure}

\begin{subfigure}{0.9\textwidth}
\centering
  \includegraphics[width=1\linewidth]{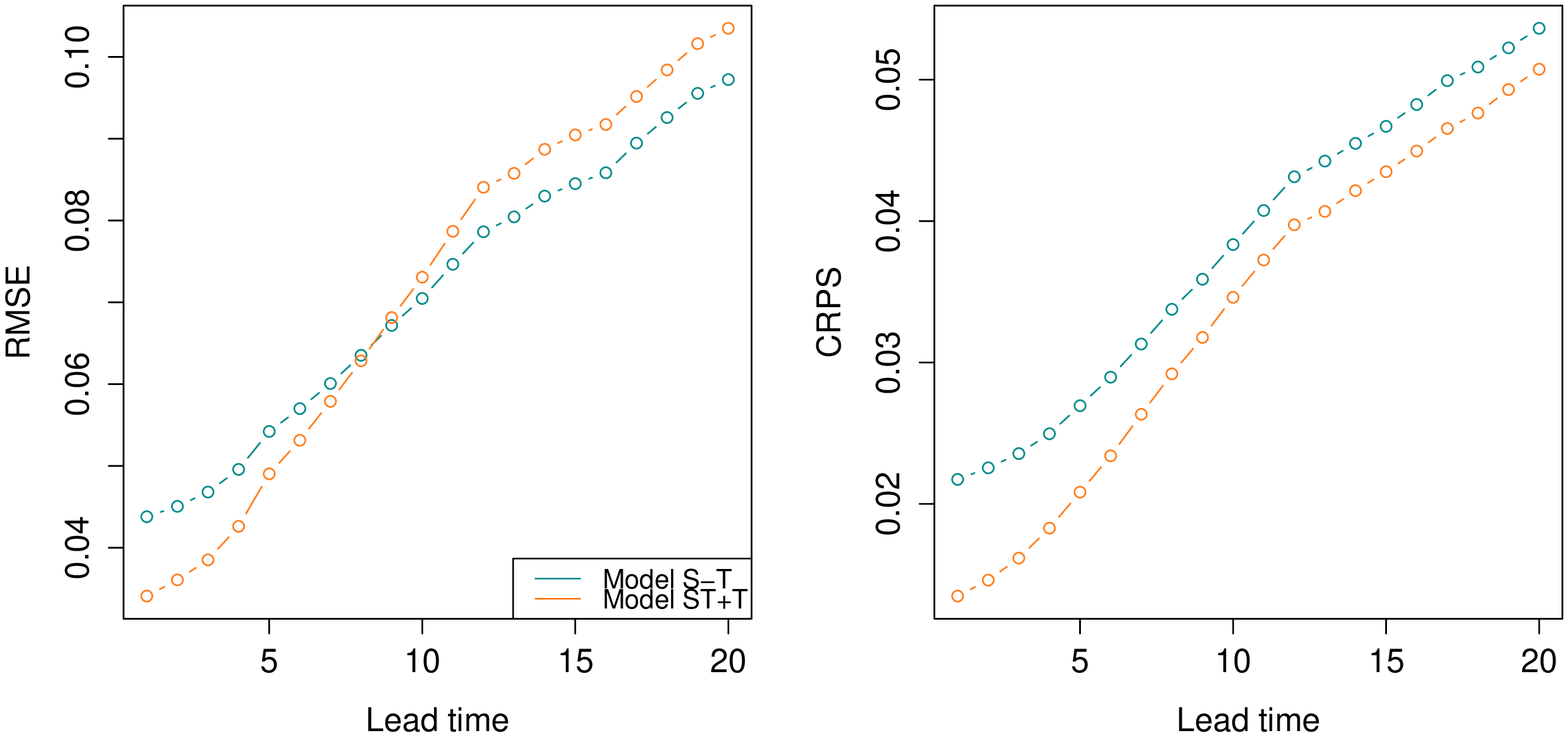}
    \caption{}
\end{subfigure}
\caption{RMSE and CRPS (as \% of nominal power) of spatially out-of-sample wind power forecasts at lead times $1, \ldots, 20$ (i.e., from 15 minutes up to 5 hours) for Model T (blue), Model S-T (green) and Model ST+T (orange). (a) Forecasts for individual wind farms. (b) Forecasts for aggregated wind farms.}
\label{spatiotemp_errors_orig}
\end{figure}

Figure \ref{spatiotemp_errors_orig} shows the out-of-sample forecast performances in terms of RMSE and mean CRPS for individual wind farms (a) and aggregated wind power (b). They are computed as the mean of the RMSE and CRPS from the 5-fold cross validations as described in Section~\ref{evaluation-scheme}. It can be seen that Model ST+T outperforms Model S-T at all lead times when predicting wind power at individual wind farms under RMSE and CPRS. When looking at aggregated out-of-sample forecasts, while for shorter lead times than 2 hours, Model S-T is better than Model ST+T in terms of RMSE, for longer horizons, Model ST+T out-performs Model S-T under the same score. In terms of CRPS, Model ST+T produces better aggregated forecasts at lead times 1-20 (i.e., from 15 minutes to 5 hours ahead).  

\begin{figure}[!htb]
\centering
\begin{subfigure}{0.9\textwidth}
\centering
  \includegraphics[width=0.8\linewidth]{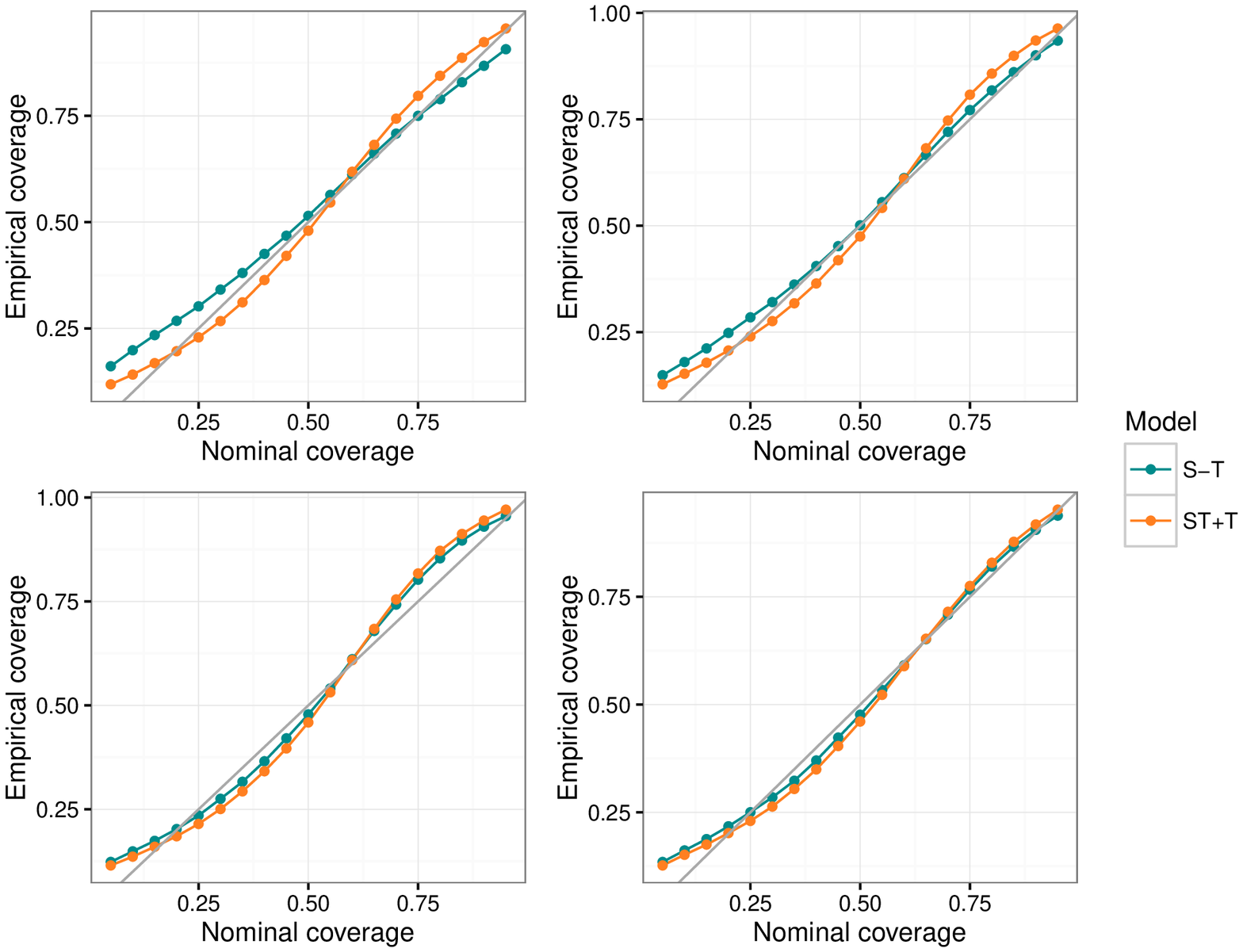}
    \caption{}
\end{subfigure}%

\begin{subfigure}{0.9\textwidth}
\centering
  \includegraphics[width=0.8\linewidth]{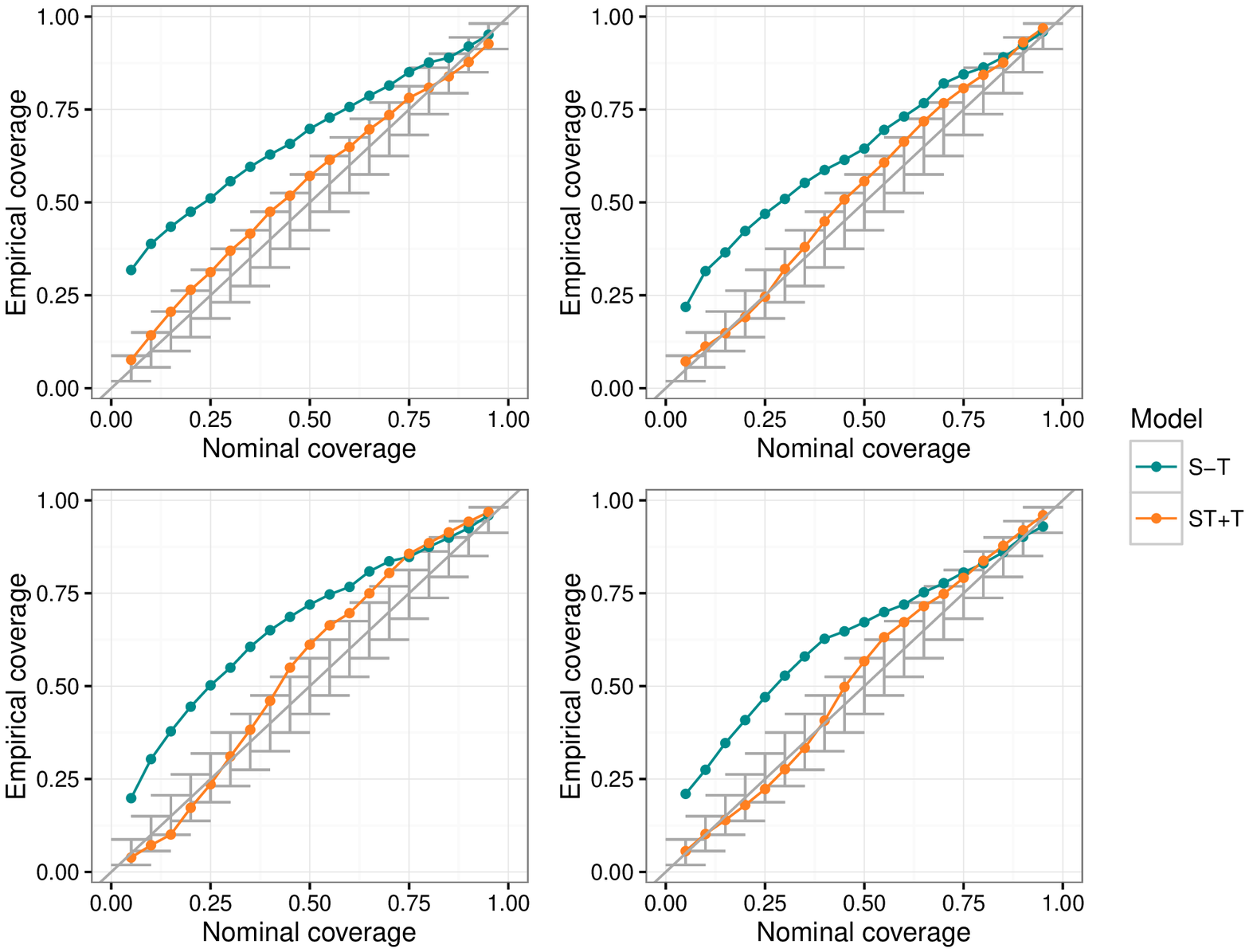}
    \caption{}
\end{subfigure}
\caption{ Reliability diagram of spatially out-of-sample wind power forecasts at lead time 1 ({\it{Top left}}), 7 ({\it{Top right}}) , 13 ({\it{Bottom left}})  and 19 ({\it{Bottom right}}). The diagrams were calculated using Model T (blue), Model S-T (green) and Model ST+T (orange). (a) Forecasts for individual wind farms. (b) Forecasts for aggregated wind farms.}
\label{spatiotemp_reliab_orig} 
\end{figure}

Reliability diagrams at lead times $h=1,7,13$ and 19 are presented in Figure \ref{spatiotemp_reliab_orig}. We observe from Figure \ref{spatiotemp_reliab_orig} (a) that Model S-T and Model ST+T provide relatively well calibrated forecast densities for individual farms. 
In terms of aggregated forecasts, we can see from Figure \ref{spatiotemp_reliab_orig} (b) that Model ST+T is calibrated, since the line is always within the consistency bars. On the other hand, aggregated forecast densities obtained with Model S-T are poorly calibrated for quantiles lower than 0.75. Indeed, 20\% of the observations are below the 5\% forecast quantile at lead times 1, 7, 13 and 19.

\subsection{Simulation study}
\label{simulation}
From the data analysis in Section~\ref{insample} and \ref{outofsample}, we see that Model ST+T is the only model among the three that produces individual and aggregated calibrated forecasts.
In this section, we simulate 200 data sets according to this model. We set the parameters equal to the estimates given by the fit of this model to one of the training data sets from our case study. More details on the evaluation scheme can be found in Section~\ref{evaluation-scheme}.

\begin{figure}[!htb]
\centering
\begin{subfigure}{0.9\textwidth}
\centering
  \includegraphics[width=1\linewidth]{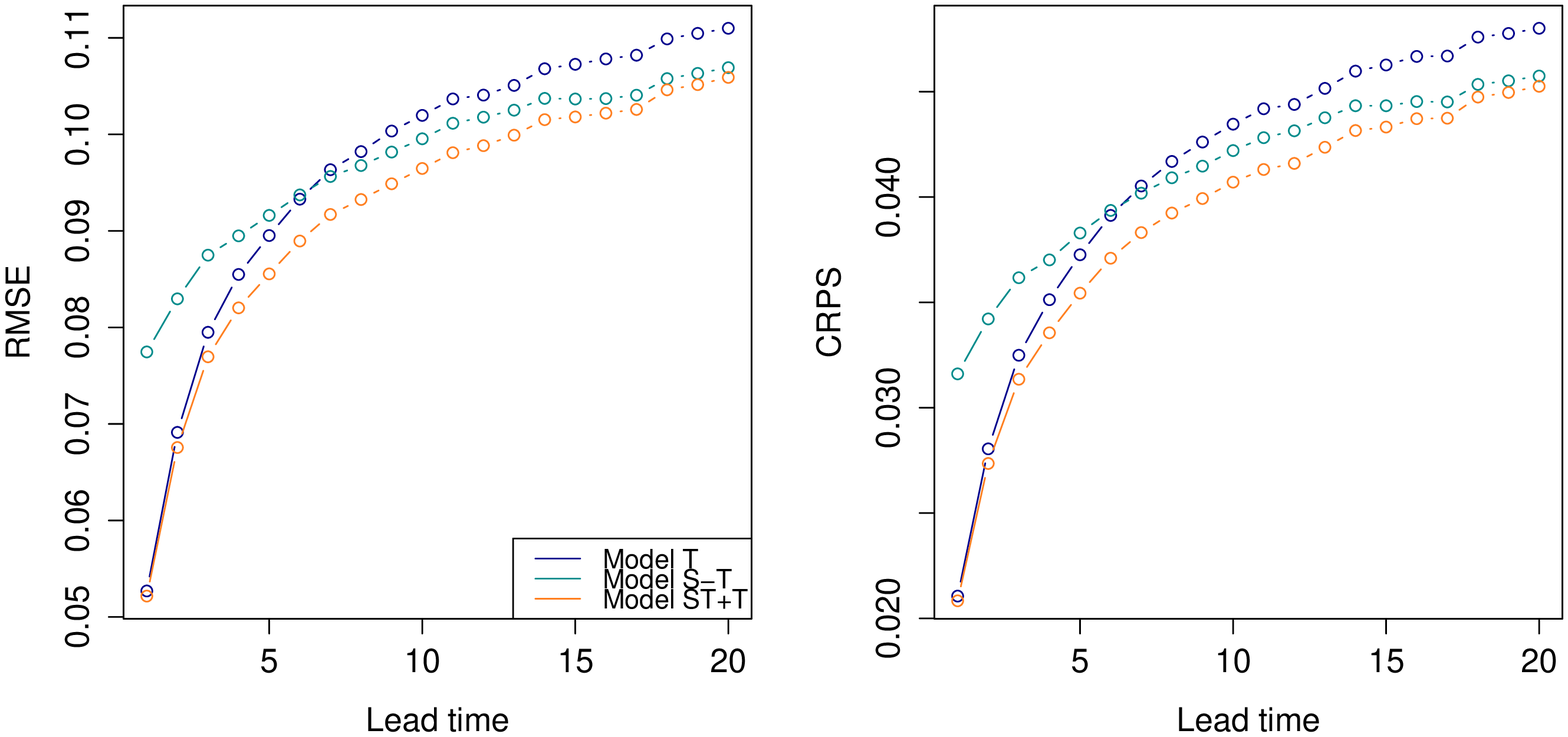}
    \caption{}
\end{subfigure}%

\begin{subfigure}{0.9\textwidth}
\centering
  \includegraphics[width=1\linewidth]{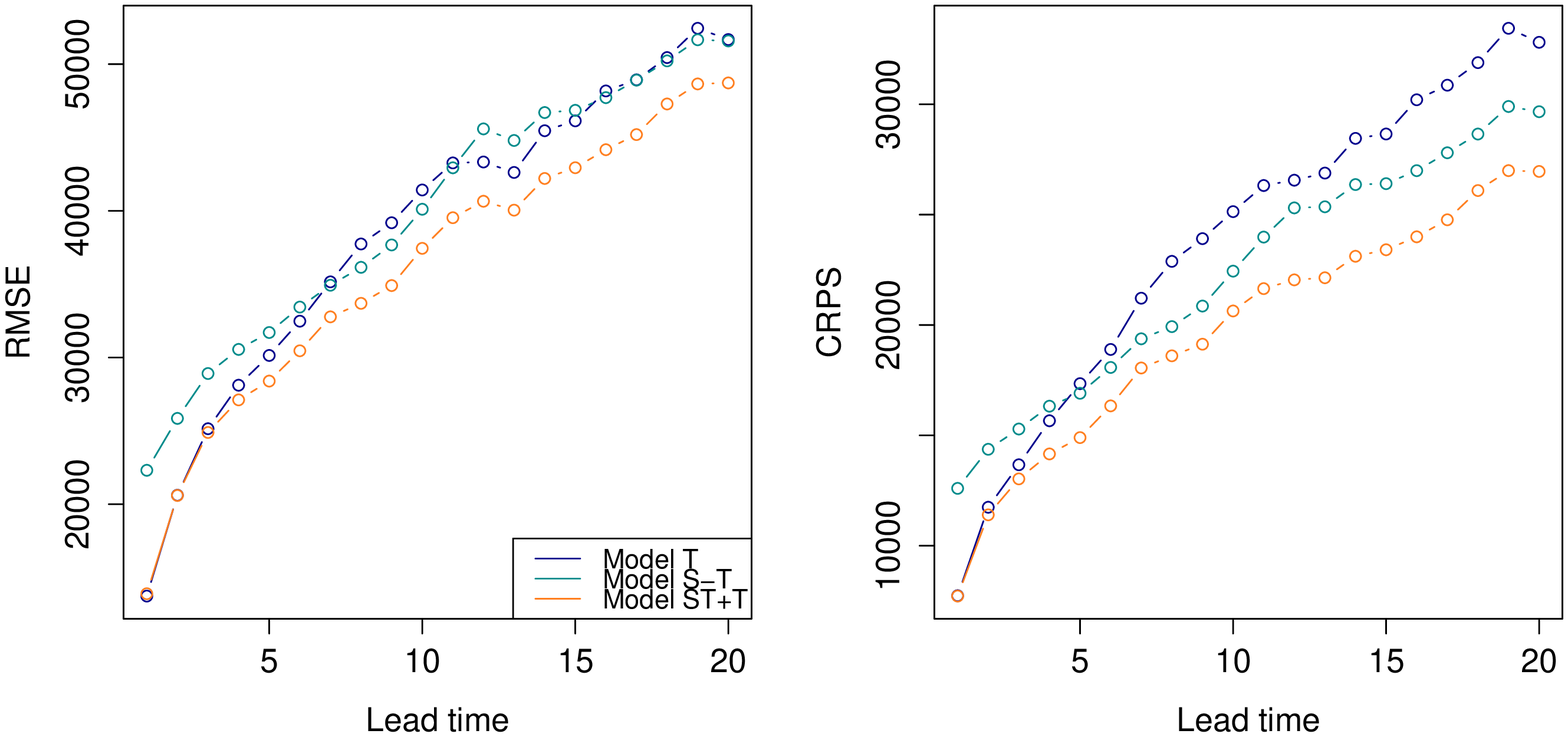}
    \caption{}
\end{subfigure}
\caption{RMSE and CRPS (as \% of nominal power) of forecasts from simulated data at lead times $1, \ldots, 20$ (i.e., from 15 minutes up to 5 hours) for Model T (blue), Model S-T (green) and Model ST+T (orange). (a) Forecasts for individual wind farms. (b) Forecasts for aggregated wind farms.}
\label{simu_errors_orig}
\end{figure}

RMSE and mean CRPS of the three different models for forecasting simulated data at lead times 1-20 are shown in Figure~\ref{simu_errors_orig}. 
According to RMSE and CRPS for individual wind farms, the three models perform similarly, while, according to CRPS for aggregated forecasts, Model ST+T out-performs Model T and Model S-T.

\begin{figure}[!htb]
\centering
\begin{subfigure}{0.9\textwidth}
\centering
  \includegraphics[width=0.8\linewidth]{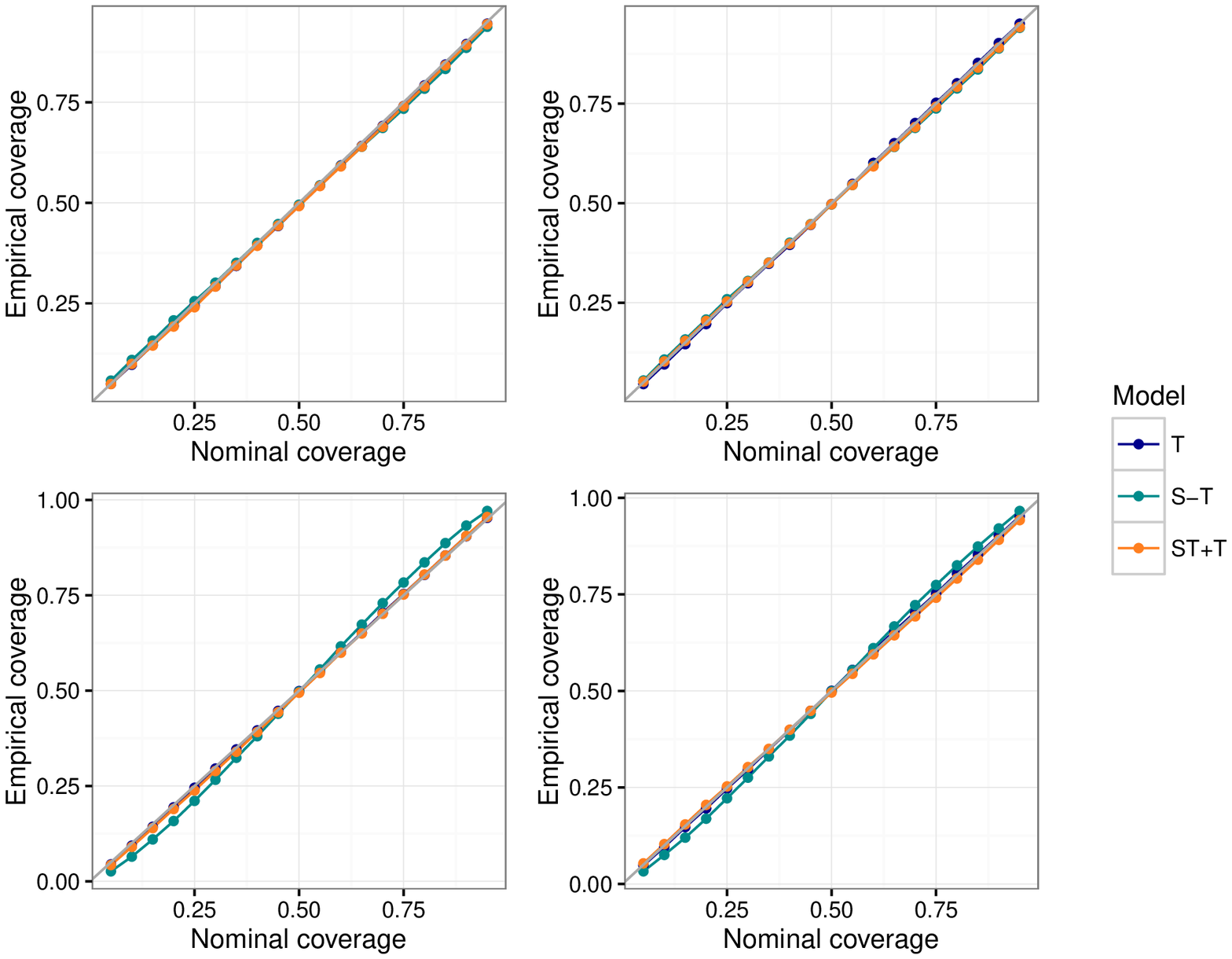}
    \caption{}
\end{subfigure}%

\begin{subfigure}{0.9\textwidth}
\centering
  \includegraphics[width=0.8\linewidth]{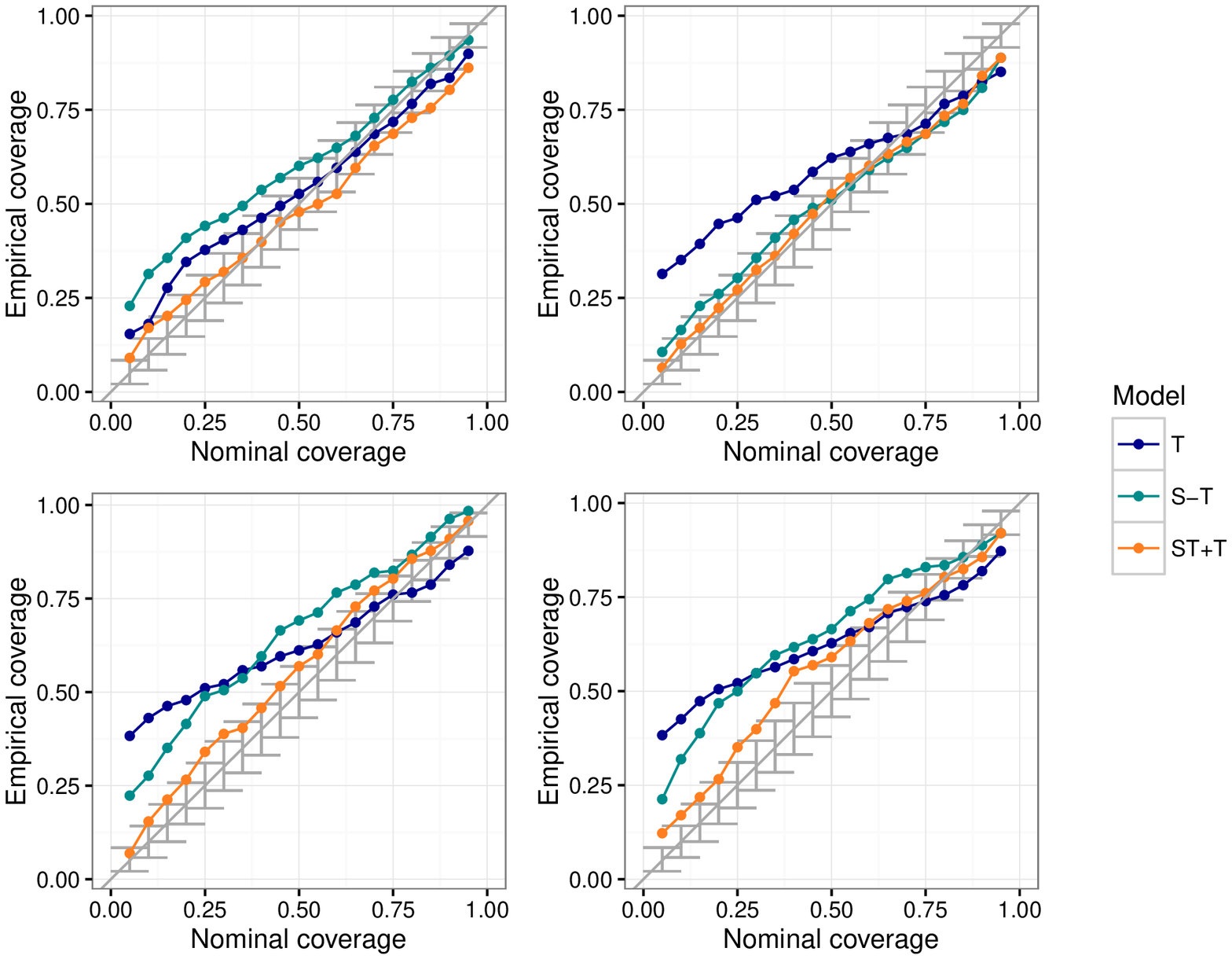}
    \caption{}
\end{subfigure}
\caption{ Reliability diagram for forecasts from simulated data at lead time 1 ({\it{Top left}}), 7 ({\it{Top right}}), 13 ({\it{Bottom left}})  and 19 ({\it{Bottom right}}). The diagrams were calculated using Model T (blue), Model S-T (green) and Model ST+T (orange). (a) Forecasts for individual wind farms. (b) Forecasts for aggregated wind farms.}
\label{simu_reliab_orig} 
\end{figure}

Results from individual and aggregated forecasts calibration are shown in Figure~\ref{simu_reliab_orig}. The forecasts from all three models are calibrated for individual wind farms, as shown in Figure~\ref{simu_reliab_orig} (a).
We observe from Figure~\ref{simu_reliab_orig} (b) that the aggregated forecasts produced by the Model ST+T are better in terms of calibration than the forecasts from the other models, which is in agreement with the results from the analysis of the aggregated Danish wind power data, as shown in Figure~\ref{reliab_orig} (b). In fact, the aggregated forecasts produced by Model ST+T are well calibrated at lead times $h=1,13,$ and 19, since the line is always inside the consistency bars.  

The simulations show that when we fit simulated data from Model ST+T using Model S-T, the spatial range $r$  (see (\ref{matern})) is underestimated. In fact, when data is generated with $r=62.1$, the first and third quartiles of the 200 estimates of this parameter from Model ST+T are 27.7 and 164.6, while with Model S-T the estimated quartiles are 25.2 and 28.0, respectively. Thus, a larger estimated spatial dependency results in a larger variance to the aggregated forecasts and makes it possible to borrow more information from close wind farms when doing out-of-sample predictions, causing the variance of a sum to increase. Hence, this explains both why the aggregated forecasts are not calibrated for Model S-T as well as why Model ST+T gives better spatially out-of-sample predictions than model S-T.

\section{Conclusions}
\label{conclusions}

In this article we have presented hierarchical spatio-temporal models for obtaining probabilistic forecasts of wind power generation at multiple locations and lead times.
We started with a time series model consisting of an autoregressive process with a location specific intercept. The results for individual probabilistic forecasts were satisfactory in terms of skill scores and reliability, however, the aggregated probabilistic forecasts were not calibrated. 
After finding the unsatisfactory results for the reliability of aggregated forecasts, we introduced two different spatio-temporal models. 
The first has a common intercept for all farms and a spatio-temporal model that varies in time with first order autoregressive dynamics and has spatially correlated innovations given by a zero mean Gaussian process with Mat\'ern covariance. The second model has a common intercept, an autoregressive process to capture the local variability and the spatio-temporal term.
To deal with the non-Gaussianity of wind power series, a parametric framework for distributional forecasts based on the logit-normal transformation was used.

In a case study, the proposed models have been used to produce probabilistic forecasts of wind power at wind farms in western Denmark from 15 minutes up to 5 hours ahead for a test period of one year. Using the SPDE approach that is implemented in the R-INLA library, we obtained fast and accurate forecasts of wind power generation at wind farms where data is available,  but also at a larger portfolio including wind farms at locations that are not included in the training set. We provided detailed analysis on the forecast performances based on appropriate metrics tailored for probabilistic forecasts. To better understand the properties of our methods, we analysed artificial data sets from a simulation study. 

Our results showed that all the proposed approaches produce calibrated short-term forecasts for individual wind farms.
However, we found that modeling spatial dependency is required to achieve calibrated aggregated probabilistic forecasts. 
Indeed, our case study  showed that spatial dependency is important for aggregated properties, and individual forecasts do not reveal this. 
Moreover, when we simulated from the spatio-temporal model containing an autoregressive term (Model ST+T), we obtained results that are in accordance with our case study, where the proposed models performed equally well for individual forecasts, while aggregated probabilistic forecasts benefit from having a spatio-temporal model with the autoregressive term. 
Model ST+T was introduced due to unsatisfactory reliability for the aggregated forecasts. Hence, evaluating aggregated forecasts can be a tool for investigating and improving models, even when spatially out-of-sample forecasts are the purpose of the modelling.
Indeed, results from spatially out-of-sample forecast performances showed that when predicting wind power at new locations that are not included in the training set, having the autoregressive term in the spatio-temporal model improved the forecast performance. 

This work was motivated by the need to produce accurate short term probabilistic forecasts at multiple wind farms and lead times, which will ultimately be applied on a national scale.
A possible extension of the models described in this work is to include weather forecast information in the linear predictor. This approach usually requires ensemble forecasts to be generated from sophisticated numerical weather prediction (NWP) models and has shown to produce reliable wind power forecasts up to 10 days ahead \cite{taylor2009wind}. 

\bibliographystyle{imsart-nameyear}
\bibliography{sample}

\end{document}